\documentclass{JHEP3}
\usepackage{amsmath,amssymb}
\newcommand{\be}{\begin{equation}}
\newcommand{\ee}{\end{equation}}
\newcommand{\beq}{\begin{eqnarray}}
\newcommand{\eeq}{\end{eqnarray}}

\newcommand{\no}{\nonumber\\}
\newcommand{\ba}{\begin{array}}
\newcommand{\ea}{\end{array}}
\def\x{\;\!}

\title{Anomalous positron excess from Lorentz-violating QED}
\author{Alexander A. Andrianov, Dom\`enec Espriu,\\Departament d'Estructura i Constituents
de la Mat\`eria and
 ICCUB Institut de Ci\`encies del Cosmos, Universitat de
Barcelona,
 Diagonal 647, 08028 Barcelona, Spain \\ E-mails: \email{andrianov@icc.ub.edu;\, espriu@ecm.ub.es}}
\author{Paola Giacconi and Roberto Soldati\\ Dipartimento di Fisica dell'Universit\`a di Bologna and
INFN - Sezione di Bologna, Via Irnerio, 46, 40126 Bologna, Italy\\ E-mails:
\email{soldati@bo.infn.it;\, giacconi@bo.infn.it}} \abstract{We entertain the idea that
a suitable background of cold (very low momentum) pseudoscalar particles or condensate,
may trigger a background that effectively generates Lorentz-invariance violation. This
\ae ther-like background induces a Chern-Simons modification of QED. Physics is
different in different frames and, in the rest frame of the pseudoscalar background,
high momentum photons can decay into pairs. The threshold for such decay depends
quadratically on the rest mass of the particles. This mechanism could explain in a
natural way why antiprotons are absent in recent cosmic ray measurements. A similar
signal could be used as a probe of pseudoscalar condensation in heavy ion collisions.}
\keywords{Space-Time Symmetries, Gamma and Cosmic Rays, QCD Phenomenology}
\preprint{UB-ECM-PF-09/19,\, ICCUB-09-218}

\begin{document}

\section{Introduction}
Recent results from the PAMELA collaboration \cite{pamela} reveal an apparent excess of
high-energy positrons in cosmic rays, which is not accompanied by a corresponding
excess in antiprotons \cite{pamela-anti}. This excess had been found previously by ATIC
\cite{atic} and PPB-BETS \cite{ppbet}. This enhancement has been confirmed recently by
the FERMI collaboration (Gamma Ray Space Telescope) \cite{fermilat} (in accordance with
previous indications from HEAT \cite{heat} and AMS-01 \cite{ams}). The latter missions
cannot separate positrons from electrons, however.

It has been pointed out that the positron excess can have a purely astrophysical
interpretation being possibly due to nearby pulsars in our galaxy  or
other astrophysical phenomena \cite{other}, but the possibility of the excess being due
to dark matter annihilation or decay is of course very interesting and has been
recently studied in some detail\cite{reviewdm}.

None of the dark matter interpretations however addresses satisfactorily the basic
puzzling question; namely why antiprotons are so conspicuously absent from the PAMELA
measurements. One would be forced to conclude that either the relevant dark matter
component is abnormally leptophilic (hadrophobic) \cite{leptoph} or the thresholds for
production of highly energetic $e^+e^-$ vs. $\bar p p$ pairs are very different due to some
unexplained effect.

In this work we address the second possibility and propose a simple mechanism by virtue
of which high-energy real photons could decay into particle-antiparticle pairs with a
momentum threshold that depends on the rest mass of the particle. We hurry to say that
this is impossible in a Lorentz-invariant theory. Indeed the mechanism is based on the
condensation of either axions \cite{ansol}, vector fields \cite{bumble} or simply
neutral pions \cite{guendelman,ae},
forming a Lorentz violating background uniform in space  (or just slowly varying in
comparison to the hard photon/high-energy positron Compton wave lengths). This
condensate should however be varying in time  (in contrast to \cite{guendelman}) for
the proposed mechanism to be possible.
We shall discuss later tentative possibilities for the origin of such effect and the
orders or magnitude involved.

It turns out that the Lorentz-violating effects \cite{carroll,liveff1,ags02,adam,liveff2} induced by this \ae ther-like
pseudoscalar density generate successive thresholds for pair productions. Obviously the
first threshold corresponds to $e^+e^-$ pairs. Photons with increasing larger values of
the momenta could eventually generate $\mu^+\mu^-$ pairs (eventually decaying to
electrons and positrons too) and for even higher energy photons $\bar p p$ pairs. A
unambiguous prediction of the model is that the successive thresholds appear for photons
whose momenta are in a relation roughly identical to the the ratio of the mass squared
of muons or protons to that one of electrons.

We shall also see that this phenomenon could also be present in heavy ion collisions if
the conditions are such that a pseudoscalar condensate may form. In fact the emergence
of electron pairs with the very specific characteristics that we will find out later
would be a novel and interesting signal of the presence of parity violation in dense
baryon matter as it has recently been suggested \cite{ae}. Another type of P-violation
in hot metastable nuclear bubbles has been proposed in \cite{kharzeev} and induced
photon instability  could be also relevant to detect it.

\section{Lorentz violation in a Chern-Simons background}
Suppose that an spatially homogeneous and isotropic background may be induced by
condensation of pseudoscalars and examine how  photons may split $\gamma
\rightarrow e^- + e^+$ via a mechanism that has a fair analogy with
the Cherenkov radiation.

The appropriate Lagrangian in the presence this background
has three pieces
\be
{\cal L} = {\cal L}_{\x{\rm INV}} + {\cal L}_{\x{\rm GF}} + {\cal L}_{\x{\rm LIV}} ;
\ee
\begin{eqnarray}
{\cal L}_{\x{\rm GF}} &=& A^\lambda(x)\,\partial_\lambda B(x) +
{\textstyle\frac12}\,\varkappa\,B^2(x) , \\
{\cal L}_{\x{\rm INV}} &=& -\,{\textstyle\frac14}\,F^{\alpha\beta}(x)F_{\alpha\beta}(x)
+ \bar\psi(x)\{\gamma^{\,\mu}[\,{\rm i}\partial_{\,\mu}-eA_{\,\mu}(x)\,]-m_e\}\,\psi(x) , \\
{\cal L}_{\x{\rm LIV}} &=& {\textstyle\frac12}\,\eta_\alpha A_\beta(x)\widetilde
F^{\,\alpha\beta}(x) ,
\end{eqnarray}
where $A_\mu$ and $\psi(x)$ stand for the photon and matter field, respectively,\\
$\widetilde
F^{\,\alpha\beta}(x)=\frac12\,\varepsilon^{\,\alpha\beta\rho\sigma}\,F_{\,\rho\sigma}(x)$
is the dual field tensor, while $B$ is the gauge-fixing auxiliary scalar field with
$\varkappa\in\mathbb R$. LIV stands for Lorentz invariance violating.

The vector $\eta_\alpha  \simeq \langle\partial_\alpha\theta\rangle \simeq
\delta_{\alpha0}\langle\dot\theta(t)\rangle $, coupled via the anomaly term to
photons, is supposedly induced by a pseudoscalar density which form the  background
in which high momentum photons, of some unspecified origin, propagate.
Similar formulae can be derived for propagating electrons and positrons
at high energies in a Lorentz violating background \cite{ags02}, but let us follow the
argumentation of KSVZ \cite{KSVZ} and neglect the kinematical distortion induced on
them.

Since this is an unfamiliar setting it is advisable to examine this theory
carefully. In momentum space
we obtain the free field equations for the LIV massive vector field and
the auxiliary scalar field
\beq
\left\{g^{\,\lambda\nu} k^2 - k^{\,\lambda}k^{\,\nu} +
{\rm i}\,\varepsilon^{\,\lambda\nu\alpha\beta}\,\eta_\alpha\,k_\beta\right\}
\tilde A_\lambda(k) + {\rm i}\,k^{\,\nu}\,\tilde B(k)=0 ,\\
{\rm i}\,k^{\,\lambda}\,\tilde A_\lambda(k) + \varkappa\,\tilde B(k)=0 .
\eeq
In the
Feynman gauge $\varkappa=1$ we get
\beq
 \tilde B(k)+{\rm i}\,k\cdot\tilde A(k)=0 ,
\label{eqmofgauge1}\\
k^2\;\tilde B(k)= k^2 \;\eta\cdot\tilde A(k)=0 ,
\label{eqmofgauge2}\\
\left\{g^{\,\lambda\nu} k^2 + {\rm
i}\,\varepsilon^{\,\lambda\nu\alpha\beta}\,\eta_\alpha\,k_\beta\right\} \tilde
A_\lambda(k)\equiv K^{\nu\lambda} \tilde A_\lambda(k) =0 . \label{eqmofgauge}
\eeq
The Levi-Civita symbol
in four dimensional Minkowski space-time is  $
\varepsilon_{\,0123} = -\,\varepsilon^{\,0123}\equiv 1 $.

We now define \be S^{\,\nu}_{\;\lambda}\ \equiv\
\varepsilon^{\,\mu\nu\alpha\beta}\,\eta_{\x\alpha}\,k_{\x\beta}\,
\varepsilon_{\,\mu\lambda\rho\sigma}\,\eta^{\x\rho}\,k^{\x\sigma}, \ee which satisfies
the following properties \be
S^{\,\nu}_{\;\lambda}\,\eta^{\x\lambda}=S^{\,\nu}_{\;\lambda}\,k^{\x\lambda}=0,\quad
S^{\,\mu\nu}\,S_{\,\nu\lambda}=\frac{S}{2}\,S^{\,\mu}_{\;\;\lambda},\quad
S=S^{\,\nu}_{\;\;\nu}= 2 [(\eta\cdot k)^2-\eta^2 k^2] . \ee Notice that for a time-like
and spatial isotropic pseudoscalar \ae ther\footnote{This type of \ae ther was
considered for the first time in \cite{carroll} but in the context of the
large-scale structure of the universe whereas here we assume its existence near some astrophysical
objects -- or in heavy ion collisions.}
$\eta^{\x\mu}=(\eta,0,0,0)$ we find $ S=2 \eta^2\,{\bf k}^2>0. $ Next, it is
convenient to introduce the two orthonormal hermitian ``projectors'' \be
P^{\,\mu\nu}_{\,\pm}\equiv \frac{S^{\,\mu\nu}}{S}\; \pm\;\frac{\rm{i}}{\sqrt{S/2}}\,
\varepsilon^{\,\mu\nu\alpha\beta}\,\eta_{\x\alpha}\,k_{\x\beta} , \ee which enjoy the
properties $\forall\,k^{\,\mu}=(\x k_0,{\bf k}\x)$ \beq
P^{\,\mu\nu}_{\,\pm}\;\eta_{\x\nu}&=&P^{\,\mu\nu}_{\,\pm}\;k_{\x\nu}=0 , \qquad
g_{\,\mu\nu}\,P^{\,\mu\nu}_{\,\pm}=1 ,
\label{projectors1}\\
P^{\,\mu\lambda}_{\,\pm}\,P_{\,\pm\,\lambda\nu}&=&
P^{\,\mu}_{\,\pm\,\nu} ,\qquad
P^{\,\mu\lambda}_{\,\pm}\,P_{\,\mp\,\lambda\nu} = 0 ,\qquad
P^{\,\mu\nu}_{+} + P^{\,\mu\nu}_{-} \ =\ \frac{2}{S}\, S^{\,\mu\nu}.
\label{projectors2}
\eeq

It follows from this that we can build up a pair of complex and space-like chiral
polarization vectors by means of a constant and space-like four vector:for example
$\epsilon_{\,\nu}=(0,1,1,1)/\sqrt3$, in such a manner that we can set
\be
\varepsilon^{\,\mu}_{\x\pm}(k)\ \equiv\ \left[\,\frac{{\bf k}^2-(\x\epsilon\cdot
k\x)^2}{2\x{\bf k}^2}\,\right]^{-\x1/2}\, P^{\,\mu\nu}_{\,\pm}\;\epsilon_{\,\nu} ,
\ee
which satisfy the orthogonality relations
\be
-\,g_{\,\mu\nu}\;\varepsilon^{\,\mu\,\ast}_{\x\pm}(k)\,\varepsilon^{\,\nu}_{\x\pm}(k)=
1 ;\qquad\quad
g_{\,\mu\nu}\,\varepsilon^{\,\mu\,\ast}_{\x\pm}(k)\,\varepsilon^{\,\nu}_{\x\mp}(k)=0 ,
\ee
as well as the closure relation
\beq
\varepsilon^{\,\mu\,\ast}_{\x+}(k)\,\varepsilon^{\,\nu}_{\x+}(k) +
\varepsilon^{\,\mu\,\ast}_{\x-}(k)\,\varepsilon^{\,\nu}_{\x-}(k)\ +\ {\rm
c.c.} = -\frac{4}{S}\,{S^{\,\mu\nu}} .\eeq

Now we are ready to find the general solution of the free field equations
(\ref{eqmofgauge}) in the Feynman gauge.  Taking into account the relations
(\ref{projectors1}) and (\ref{projectors2}) we readily obtain \be K^{\,\mu}_{\;\;\nu} =
\delta^\mu_{\,\nu} k^2 + \sqrt{\frac{S}{2}} (P_{+\,\nu}^\mu - P_{-\,\nu}^\mu). \ee
Therefore \be K^{\,\mu}_{\;\;\nu} \varepsilon^\nu_\pm(k) = \left(k^2 \pm
\sqrt{\frac{S}{2}}\right) \varepsilon^\mu_\pm(k), \ee which shows that they are
solutions of the equations of motion iff \be k^{\,\mu}_{\,\pm}=(\omega_\pm({\bf
k})\, , \,{\bf k})\qquad\quad \omega_\pm({\bf k})=\displaystyle \sqrt{{\bf k}^2
\pm\eta\,|\x{\bf k}\x|} . \ee Evidently for ``-'' polarization the photon energy at
sufficiently low momenta, $|{\bf k}|< \eta$, becomes imaginary signifying the instability
of photon vacuum \cite{ansol,carroll,ags02,adam}.

For strictly vanishing photon mass the instability affecting the ``-'' polarization
sets up already for zero-momentum photons. However in-medium photons do acquire a mass
without breaking any fundamental gauge principle, so we will take $m_\gamma\neq 0$. The
instability then sets up for $\eta > 2 m_\gamma$ (see \cite{aacgs06}) and whether the
vacuum instability it is relevant or not depends on the relative values of $m_\gamma$
and $\eta$. However in this work we are interested in high-energy phenomena when photon
wave vectors are much larger than the scale set up by $\eta$  (in fact assuming that
the very concept of a spatially {\it\emph constant} pseudoscalar background is valid at
microscopic scales and makes sense for sufficiently high photon frequencies ).
Accordingly we neglect this subtlety and refer the reader for its treatment to
\cite{aacgs06}.

It is also important to realize that for ``-'' photons, even after having introduced a
mass $m_\gamma > \eta/2$ the condition $k^2 \geq 0$, which is necessary for genuine
causal propagation, holds only and only if the spatial momentum ${\bf k}$ stands below
the momentum cutoff $\Lambda_{\,\gamma}\,,$ i.e., it belongs to the causality/stability
region \be |\x{\bf k}\x|<\frac{m^{\x2}_{\x\gamma}}{\eta}\equiv\Lambda_{\,\gamma}
.\label{stability} \ee Above this bound, photons of both chiralities would obviously
decay (via the box diagram) to three photons of negative chiralities which in turn
could also decay and so on until they are completely red-shifted. The decay process is
a slow one, nevertheless being of order $\alpha^4$. It happens because photons of
negative chiralities obtain an effective mass of tachyonic type  having nevertheless
group velocities less than the conventional speed of light \cite{ags02} .

We introduce two further orthonormal polarization four-vectors, i.e., the temporal and
longitudinal polarization vectors, respectively
\beq
\varepsilon^{\,\mu}_{\,T}\x(k)\;&\equiv&\; \frac{{\rm i}\x
k^{\,\mu}}{\sqrt{\,k^2}}\qquad\qquad
(\,k^2>0\,) ,\\
\varepsilon^{\,\mu}_{\,L}\x(k)\;&\equiv&\;(\x k^2\x\mbox{\tt D}\,)^{-\,1/2}\left(
k^2\,\eta^{\,\mu} - k^{\,\mu}\,\eta\cdot k\,\right)
\qquad\qquad
(\,k^2>0\,) ,
\eeq
which fulfill by construction
\be
g_{\,\mu\nu}\;\varepsilon^{\,\mu\,\ast}_{\,T}\x(k)\,\varepsilon^{\,\nu}_{\,T}\x(k)=
-\;g_{\,\mu\nu}\;\varepsilon^{\,\mu}_{\,L}\x(k)\,\varepsilon^{\,\nu}_{\,L}\x(k)= 1 ;
\qquad
g_{\,\mu\nu}\,\varepsilon^{\,\mu}_{\,T}\x(k)\,\varepsilon^{\,\nu}_{\,L}\x(k) = 0 .
\ee
Thus we have at our disposal $\forall\,k^{\,\mu}$ with $k^2>0$ a complete
orthonormal set of four polarization four vectors : namely $\varepsilon_A^\mu(k)$ with
$A=T,L,+,-$. They satisfy
\be
g_{\,\mu\nu}\;\varepsilon^{\,\mu\,\ast}_{\,A}\x(k)\,
\varepsilon^{\,\nu}_{\,B}\x(k)\,=\,g_{\,AB} ;\qquad\quad
g^{\,AB}\,\varepsilon^{\,\mu\,\ast}_{\,A}\x(k)\,\varepsilon^{\,\nu}_{\,B}\x(k)\,=\,g^{\,\mu\nu},
\label{orthoclosure}
\ee
where $g^{AB}= diag (1,-1,-1,-1)$.

Then for massive photons the physical subspace consists of the three polarizations
$A=\pm, L$, but one of the two chiral transverse states with complex polarization
vectors $\varepsilon^{\,\nu}_{\,\pm}(k_{\x\pm})$, namely
$\varepsilon^{\,\nu}_{\,-}(k_-)$ exists only iff $|\x{\bf
k}\x|<\Lambda_{\,\gamma}\,\Leftrightarrow\,k_{\x-}^{\x2}>0\,.$ If we take $m_\gamma=0$
this helicity state becomes superluminal for all values of the momenta and produces a
sort of Cherenkov radiation, gradually splitting into three  photons with negative
polarizations (see \cite{klink} for similar arguments but for space-like background
vectors). We have to stress that, kinematically, the high-energy photon with positive
polarization can {\em also} undergo splitting into the negative polarization photons.
Both splittings are kinematically allowed as it can be easily read out from the
inequality for the forward decay (we neglect here the photon mass), \be \omega_\pm
({\bf k})=\displaystyle \sqrt{{\bf k}^2 \pm\eta\,|\x{\bf k}\x|} > 3\,
\omega_-(\frac{{\bf k}}{3}) . \ee Thus if the phenomenon of positron excess is
accounted for by the instability of photons in a pseudoscalar background an
accompanying effect might be the suppression of high-energy $\gamma$ rays from the same
region, depending on the value of the effective photon mass, bearing in mind that this
process is anyway a one-loop effect and the threshold is  $\sim m_\gamma^2/\eta$
(perhaps the results reported in \cite{fan} might be a hint of this phenomenon).
In addition, there is the
possibility of ``radiative'' LIV decays $e^-\to e^- \gamma$; the momentum threshold
being $|{\bf k}| > m_\gamma m_e / \eta$. This effect will change the energy spectrum of
the $e^+ e^-$ pair produced in LIV $\gamma \to e^+ e^-$ decays, but it is suppressed by
a power of $\alpha$ and the cross-section must be proportional to $\eta$ too.

\section{Decay amplitudes}

We are now ready to derive
the lowest order decay amplitude for the process $\gamma\to e^+ e^-$. The Feynman rules are
formally the usual ones except for the addition of thresholds such as the one implied by
(\ref{stability}).
\be
|\x\mathcal M(k,p,q)\x|^{\x2} = 2\pi\,\alpha\;
g_{\,\rho\sigma}\,g_{\,\mu\nu}\,
\varepsilon_{\,A}^{\,\mu}({\bf k})\,\varepsilon_{\,B}^{\,\rho\,\ast}({\bf k})
\times
\bar u_{\,r}(p)\,\gamma^{\x\nu}\,v_{\,s}(q)\;\bar v_{\,s}(q)\,\gamma^{\x\sigma}\,u_{\,r}(p)\ .
\ee
Let us now determine the thresholds for the different polarizations
 involved in the decay process.
For the longitudinal polarization $A=L$ the dispersion relation is $k_0^2={\bf
k}^2+m_{\x\gamma}^{\x2}$ so that the energy-momentum conservation forbids the decay
process for $m_{\x\gamma}\ll m_{\x e}\,.$ Conversely, for the transverse chiral
polarizations $A=\pm$ we find \beq \omega_\pm{\,(\bf k})=\displaystyle \sqrt{{\bf
k}^2+m_{\,\gamma}^{\x2}\pm\eta\,|\x{\bf k}\x|} =\sqrt{{\bf p}^2+m_{\x e}^{\x2}} +
\sqrt{({\bf k}-{\bf p})^2+m_{\x e}^{\x2}} \equiv E({\bf p}) + E({\bf
q}) ,\eeq with ${\bf q}={\bf k} - {\bf p}$ . From the energy-momentum conservation we get
\beq k_{\x\pm}^{\x2}=m_{\,\gamma}^{\x2}\pm\eta\,|\x{\bf k}\x| \simeq \pm\eta\,|\x{\bf
k}\x| =2m_{\,e}^{\x2}+2 E({\bf p}) E({\bf q})-2{\bf p}\cdot{\bf q}, \label{encons} \eeq since
$m_{\,\gamma}\ll m_{\,e}$ .  It is evident that photons with a negative chiral
polarization $A=\,(-)\,$ cannot decay as they are tachyonic-like in such a background,
while photons of positive chiral polarization $A=\,(+)\,$ do undergo the decay iff the
photon momentum is above the threshold \be |\x{\bf
k}\x|\ge\frac{4m_{\,e}^{\x2}}{\eta}\equiv k_{\,\rm th} . \ee

The calculation of  the total decay width is quite standard: \beq \Gamma_{\,+}\x({\bf
k}) &=& \frac{1}{2\x\omega_+({\bf k})} \int\frac{{\rm d}\x\bf p}{(2\pi)^3\x2\x E({\bf
p})} \int\frac{{\rm d}\x\bf q}{(2\pi)^3\x2\x E({\bf q})}\,
(2\pi)^4\,\delta^{(4)}(k-p-q) \no&&\times\sum_{r\x,\x s\,=\,1,2}\x|\mathcal M_{\,r\x
s\x+}\x(k,p,q)\x|^{\x2} \no&=& \frac{1}{32 \pi^2 \x \omega_+({\bf k})} \int\frac{{\rm
d}\x\bf p}{E({\bf p})\x E({\bf q})}\, \delta \Big(\omega_+({\bf k})- E({\bf p})- E({\bf
k - p})\Big) \no&&\times\sum_{r\x,\x s\,=\,1,2}|\x\mathcal M_{\,r\x
s\x+}\x(k,p,k-p)\x|^{\x2} .\label{phasein}\eeq
 From the equality \eqref{encons} in the form \beq k_+^2 = \eta\,|\x{\bf
k}\x|  =  2E({\bf p})\,[\,E({\bf p}) + E({\bf k}-{\bf p})\,] - 2\x|\x{\bf
k}\x|\,|\x{\bf p}\x|\,\cos\theta ,\eeq we readily get with $k=|\x{\bf k}\x|$ and
$p=|\x{\bf p}\x|$ the following condition for the energy delta function to be satisfied
\beq {\frac12}\,\eta\,k  =\displaystyle\sqrt{(p^{\x2}+m^{\x2}_{\,e})(k^{\x2}+k\x\eta)}
- k\x p\x\cos\theta = p_\mu k^{\mu}_{\x+}, \label{equa}  \eeq which gives the physical
values of electron/positron momenta, \be p_\pm = \frac{\eta \cos\theta \pm 2m_e
\sqrt{\Big(1+\frac{\eta}{k}\Big)\Big(\frac{\eta^2}{4m^2_e}\Big(1-\frac{4m_e^2}{\eta
k}\Big)- \sin^2\theta\Big)}}{2\Big(\sin^2\theta + \frac{\eta}{k}\Big)} ,\ee and
requires the following inequality to hold \be
\sin^{\x2}\theta\leq\frac{\eta^{\x2}}{4m^{\x2}_{e}}\Big(1-\frac{4m_e^2}{\eta k}\Big) .
\ee As we see, the emitted pair is produced inside a narrow forward cone with an angle
that, typically, should be small as we expect $\eta << m_e$; that is $\theta_{\,\rm
max} < \eta/{2m_{\,e}}$ .

For the adjacent momentum $\bf q = k - p$ we define $q = |{\bf k - p}|,\, {\bf q\cdot
k} = q k \cos\tilde\theta$ and find the relation
 \be q \sin\tilde\theta = - p \sin\theta, \ee wherefrom and from eq.\eqref{equa}
it can be obtained that the two angles
are similarly small but the two momenta are complementary. For instance, if $k \gg k_{\,\rm
th}$  for the lowest
limit $p_-\simeq \frac{m^2_e}{\eta}$ one finds $q_+ \simeq k - \frac{m^2_e}{\eta}$ and
vice versa.

The angular integration in the phase space integral \eqref{phasein} resolves the delta-function in a nontrivial way when
\be
\cos\theta = - \frac{\eta}{2p} + \displaystyle\sqrt{\Big(1+\frac{\eta}{k}\Big)\Big(1+\frac{m_e^2}{p^2}\Big)} \leq 1, \label{cos}
\ee
that  bounds the momentum  \be \mbox{\rm min}\, p_{-}\simeq \frac{m_e^2}{\eta} \leq p \leq \mbox{\rm max}\, p_{+}\simeq k - \frac{m_e^2}{\eta} ,\ee
for $k \gg k_{\,\rm
th}$ .  Accordingly the phase space integral can be evaluated
\beq
\frac{1}{32\pi^2 \omega_+({\bf k})} \int
\frac{{\rm d}\x\bf p}{\x E({\bf p})\x E({\bf k- p})}\,
\delta\Big(\omega_+({\bf k})+ E({\bf p})+ E({\bf k - p})\Big) \simeq \frac{1}{16 \pi k^2} \int\limits^{k - \frac{m_e^2}{\eta}}_{\frac{m_e^2}{\eta}} {\rm d}\x p ,
\eeq
for $\cos\theta$ obeying eq.\eqref{cos}.

Finally, setting $m_\gamma=0$, the squared decay amplitude for a photon of positive
chiral polarization is
\beq &&\sum_{r\x,\x s\,=\,1,2}|\x\mathcal M_{\,r\x
s\x+}\x(k,p,q)\x|^{\x2} = 16\x\pi\,\alpha\; \theta\left(\,|\x{\bf k}\x|-k_{\,\rm
th}\x\right) \times \left[\,2\x p_\mu\,p_\rho\,P^{\,\mu\rho}_{\,+}(k_+) + p\cdot
k_{\x+}\,\right]\no
&&= 16\x\pi\x\alpha\, \left\{ m^2_e - \frac14 \eta^2  + \frac{\eta}{2k}\Big( E^2({\bf p})+ E^2({\bf k - p}) \Big)\right\}\no
&& \simeq \frac{8\x\pi\x\alpha\x \eta}{k}\Big( 2p^2 +k^2 -2 k p \Big)  \eeq
for sufficiently large $k \gg k_{\,\rm
th}$ . Eventually the integration over phase space entails
\be \Gamma_{\,+}\;=\tau_{\,+}^{\x-1}\;\simeq\;
\frac{\alpha\eta}{3}, \ee
which is essentially constant for high-energy photons.

\section{Physical scenarios}

We have seen that the presence of an \ae ther-like time-dependent pseudoscalar
background leads to the possibility of rather exotic phenomena\footnote{For analogous
processes triggered by a {\it\emph space-like} CS vector, see the analysis in
\cite{klink}.} such as $\gamma\to e^+ e^-$ (and  $e\to \gamma e,\ \gamma \to
\gamma\gamma\gamma$ controlled by an effective photon mass). For a LIV photon we have
seen that this is a physical state only for a given polarization governed by the sign
of the background vector $\eta_\mu$ whereas photons of the opposite polarization are
"tachyonic" or superluminal in the phase velocity. Furthermore the decay of physical
photons can take place (due to energy-momentum conservation) for photons of
sufficiently high 3-momentum. This leads to the possibility of successive thresholds
for the production of progressively more massive $\bar f f $ pairs. The key inequality
is \be \vert {\bf k} \vert > \frac{4 m_f^2}{\eta}. \ee

The accompanying processes
$e\to \gamma e,\, \gamma \to \gamma\gamma\gamma$ are controlled by an effective photon mass and by an oscillation frequency of the axion background $\eta$. They involve "tachyonic" photons as final
states and, although they are rare, they potentially lead to a red shift in
electron/positron and photon spectra as well as to the dominance of a particular photon
polarization if $\eta$ has a definite sign. But most plausibly $\eta$ is slowly oscillating as being the derivative of an axion-like background which cannot grow up to extremely large values. Then if the phenomenon takes place in a sufficiently large volume one cannot register a definite polarization.

The decay width of the process $e\to \gamma e$ is
expected to be of the same order $\sim \alpha \eta$ as the photon decay due to crossing symmetry. But the final photon obtains a tachyonic effective mass and is not involved further on into $e^+ e^-$ pair creation (but rather decays in 3-photon states). Thus this decay enforces a red shift in the electron spectrum, in principle observable by Fermi-LAT \cite{fermilat}. In particular, the absence or softening of the electron-positron excess \cite{fermilat} in the interval of $300\div800$ GeV might be just accounted for by this red shift. Thus such a phenomenon may perhaps point
out to a mechanism of anomalous $e^+ e^-$ pair creation as a main source for
explanation of the PAMELA data. In turn the decay width of photon splitting $\gamma \to \gamma\gamma\gamma$ is estimated to be much smaller, proportional to $\alpha^2 \eta^2/m_f$ and is not expected to make an essential red shift of the photon spectrum. More accurate calculations have not been done yet for time-like $\eta$ and this work is in progress.

Let us now discuss possible physical situations where this phenomenon might happen. Let us begin
by discussion of heavy ion collisions and nuclear matter at high densities. It has
been derived in \cite{ae} that a phase where parity is spontaneously broken may
exist for baryonic densities corresponding to 3 to 8 times the usual nuclear density.
Such a state could be produced
in heavy ion collisions in conditions which are expected in the experiment CMB \cite{cmb}
 at the FAIR facility or at the planned NICA accelerator \cite{nica}. As the density in
the center-of-mass frame of colliding ions is growing in time of collision one
anticipates a time-dependent neutral pion condensate with a nearly constant derivative.
The latter just produces a modification of QED and photon instability as described
above. In this case the natural scale for $\eta$ is \be \eta \sim
\frac{\alpha}{2 \pi}\frac{\dot \rho}{f_\pi} \frac{\partial \langle
\Pi\rangle}{\partial \rho}, \ee where $f_{\pi}$ would be the in-medium pion decay
constant (vacuum value: $f_\pi \sim 100$ MeV), $\rho$ is the density and $\langle
\Pi\rangle$ is the value of the parity-breaking condensate. Although the actual
magnitude depends on interaction details (such as $\dot\rho$, which in turn depends on
the hadronic matter compressibility) a natural expectation would be $\eta \sim 1$ keV
implying that the threshold for anomalous LIV $e^+e^-$ pair production $k_{\,\rm
th}$ is naturally in the GeV
region and the photon decay width $\Gamma_{\,+}$ is of order 1 eV for $k \gg k_{\,\rm
th}$ . As for hard photons, they are abundant due to interactions at the parton level
with energies in the GeV region. So anomalous LIV $e^+e^-$ pair production is a realistic
possibility and a potential way to probe hadronic matter in extreme conditions. It is
remarkable that the generation of photon mass is also predicted in the presence of a
neutral pion condensate \cite{guendelman}.

In an astrophysical context, it has been seen by the PAMELA mission that there is an
excess of positrons starting at around 10 GeV . So we assume $k_{\rm th}$ to be
approximately 10 GeV. An obvious possibility would be to consider the relic cold axion
\cite{axion} density. Then, for a light axion with $f_a \sim 10^{11}$ GeV, a
calculation analogous to the just presented would lead to \be \eta \sim
\frac{\alpha}{2\pi}\frac{1}{f_a} \sqrt{m_a n_a}, \ee where $m_a$ is the axion mass
$< 1$eV and $n_a$ the number of cold axions per unit volume. Thus this possibility to
explain the PAMELA excess is totally excluded as it would require $n_a$ many orders of
magnitude larger than the current cosmological bound. Somewhat lower bounds on $f_a \leq 10^{10}$ GeV
have been obtained with different helioscopes \cite{helio} to register solar/terrestrial
axion-like particles \footnote{Some time ago, in the PVLAS experiments, much larger values for axion-like particle mass and much lower values for $f_a$ have been announced. But later, after substantial improvement of the experimental technique the birefringence effect disappeared \cite{pvlas} and the bounds are conservatively given by the CAST observations \cite{cast}.}(see the reviews in \cite{axion} ) but yet its value seems to be too low to trigger a visible photon instability within the solar system. Thus the electron-positron excess must have its origin far from the solar system.

Another possibility to consider is the apparent existing window for more
massive axions \cite{massax} which
corresponds to $f_a\sim 10^5 $ GeV. As we have said to generate anomalous LIV pair in the GeV
region one needs $\eta$ to be in the KeV region. This would require still very large
axion densities, perhaps attainable only in axion stars. Finally, a very fast varying
time dependent axion condensate could provide the seed for LIV anomalous pair production, but
we do not have a concrete mechanism to propose.

If cold axions \cite{axion} constitute the essential component of dark matter, we can
estimate the value of $\eta$ to be $\eta \sim 10^{-29}$ eV . This may seem a tiny
value, and indeed it is, but if we take our previous results at face value and use the
upper limit for an effective photon mass in the range of $10^{-18}$ eV \cite{PDG}, we
see that the threshold for the anomalous LIV "radiative" process $e^- \to e^- \gamma$
is $10^8$ GeV or lower, while the one for $p \to p \gamma$ is $10^{11}$ GeV . Both
should be relevant for high-energy cosmic rays, acting as an additional suppression on
top of the usual GZK effect.

There may be more exotic causes to generate an axial-vector condensate. Condensation of
vector ``bumblebee`` fields
in certain non-renormalizable models \cite{bumble} and of gradients of (nearly) massless axions
due to Coleman-Weinberg
mechanism \cite{ansol}. Unfortunately in those approaches the identification of
dark condensates cannot be easily done on terms of particle physics.

It is anyhow clear that to observe anomalous thresholds in the GeV region the natural scale is provided by strongly interacting pseudoscalars. The experimental signal would be rather unmistakable: the produced $e^+e^-$ pair flies away in a very narrow cone and, of course, the process has a threshold that has nothing to do with the usual $\gamma\gamma \to e^+e^-$ one.

\section{Conclusions}

In this paper we have investigated the phenomenology of Lorentz-violating QED,
described by an additional Chern-Simons term including a time-like, but translational
invariant, background. This \ae ther-like background could be provided by any type of
cold pseudoscalars. We have also explored the possibility that this LIV modification of
QED may trigger anomalous $e^+ e^-$ pair production. In fact it is totally natural in
this type of models to have very different thresholds, thus explaining in a simple and
natural way why $\bar p p$ pairs are conspicuously absent in some astrophysical
phenomena.

Axions or
axion-like particles are obvious candidates for this pseudoscalar background, but if the
anomalous pair production is to take place around 10 GeV this requires an absurdly large density
of relic cold axions. A more like scenario is that pseudoscalar condensation due to strong
interactions may give some visible effects.

\acknowledgments We thank the hospitality offered to two of us by the Centro de
Ciencias de Benasque Pedro Pascual where this work was finished. We acknowledge
financial support from projects FPA2007-66665-C02-01, 2009SGR502 and Consolider CPAN
CSD2007-00042 as well as the MICINN-INFN exchange agreement FPA2008-03836-E. the work
of A.A. is partially supported by Grant RFBR 09-02-00073 and Program RNP2009-1575. We
thank F. Lizzi for discussions.


\bigskip
\begin{thebibliography}{99}
\bibitem{pamela} O. Adriani et al., \nature{458}{2009}{607}; \arXivid{0810.4995} [astro-ph].
\bibitem{pamela-anti} O. Adriani et al., \arXivid{0810.4994} [astro-ph].
\bibitem{atic} J. Chang et al., \nature{456}{2008}{362} .
\bibitem{ppbet} S. Torii et al. [PPB-BETS Collaboration], \arXivid{0809.0760} [astro-ph].
\bibitem{fermilat} A. A. Abdo et.al., [Fermi LAT Collaboration], \prl{102}{2009}{181101}.
\bibitem{heat} S. W. Barwick et al. [HEAT Collaboration], \apj{482}{1997}{L191} [\astroph{9703192}];\\
 J. J. Beatty et al., \prl{93}{2004}{241102} [\astroph{0412230}].
\bibitem{ams} M. Aguilar et al. [AMS-01 Collaboration], \plb{646}{2007}{145} [\astroph{0703154}].
\bibitem{other} D. Hooper, P. Blasi and P. D. Serpico, \newjournal{JCAP}{JCAPA}{0901}{2009}{025}
[\arXivid{0810.1527} [astro-ph]] ;\\ S. Profumo, \arXivid{0812.4457}
[astro-ph];\\ H. Yuksel, M. D. Kistler and T. Stanev, \prl{103}{2009}{051101};\\ N. J. Shaviv, E. Nakar
and T. Piran, \arXivid{0902.0376} [astro-ph.HE];\\ S. Dado and
A. Dar, \arXivid{0903.0165} [astro-ph.HE];\\ D. Malyshev, I. Cholis
and J. Gelfand, \arXivid{0903.1310} [astro-ph.HE];\\ Y. Fujita,
K. Kohri, R. Yamazaki and K. Ioka, \arXivid{0903.5298}
[astro-ph.HE];\\ V. Barger, Y. Gao, W. Y. Keung, D. Marfatia
and G. Shaughnessy, \arXivid{0904.2001} [hep-ph];\\ D. Grasso {\it et al.}  [FERMI-LAT Collaboration],
\arXivid{0905.0636} [astro-ph.HE];\\
 P. Mertsch and S. Sarkar,
\arXivid{0905.3152} [astro-ph.HE].
\bibitem{reviewdm} M. Cirelli, M. Kadastik, M. Raidal and A. Strumia, \npb{813}{2009}{1};\\ N. Arkani-Hamed, D. P. Finkbeiner, T. Slatyer and N. Weiner, \prd{79}{2009}{015014};\\  M. Pospelov and A. Ritz, \plb{671}{2009}{391};\\ P. F. Yin, Q. Yuan, J. Liu, J. Zhang, X. J. Bi, S. H. Zhu and X. M. Zhang, \prd{79}{2009}{023512};\\ D. Hooper, A. Stebbins and K. M. Zurek, \prd{79}{2009}{103513}
;\\
K. Ishiwata, S. Matsumoto and T. Moroi, \plb{675}{2009}{446};\\
 F. Chen, J. M. Cline and A. R. Frey,\prd{79}{2009}{063530}
;\\ A. Arvanitaki, S. Dimopoulos,
S. Dubovsky, P. W. Graham, R. Harnik and S. Rajendran,
\arXivid{0904.2789} [hep-ph];\\ M. Regis and P. Ullio, \arXivid{0904.4645} [astro-ph.GA];\\ X. Calmet
and S. K. Majee, \arXivid{0905.0956} [hep-ph];\\ S. Shirai,
F. Takahashi and T. T. Yanagida, \arXivid{0905.0388}
[hep-ph];\\ P. Meade, M. Papucci, A. Strumia and T. Volansky,
 \arXivid{0905.0480} [hep-ph];\\
C. H. Chen, C. Q. Geng and D. V. Zhuridov,
\arXivid{0905.0652} [hep-ph];\\ K. Hamaguchi, K. Nakaji
and E. Nakamura, \arXivid{0905.1574} [hep-ph];\\  T. Delahaye et al., \arXivid{0905.2144} [hep-ph];\\ N. Okada and T. Yamada,
\arXivid{0905.2801} [hep-ph];\\ H. Fukuoka, J. Kubo and D. Suematsu,
\arXivid{0905.2847} [hep-ph];\\ Y. Bai, M. Carena and
J. Lykken, \arXivid{0905.2964} [hep-ph];\\ S. Shirai, F. Takahashi
and T. T. Yanagida, \arXivid{0905.3235} [hep-ph];\\ C. H. Chen,
\arXivid{0905.3425} [hep-ph];\\ J.Mardon, Y. Nomura and J. Thaler,
\arXivid{0905.3749} [hep-ph] .
\bibitem{leptoph} P.J. Fox and E. Poppitz,
\prd{79}{2009}{083528};\\ B. Kyae, \arXivid{0902.0071}
[hep-ph];\\ X. J. Bi, X. G. He and Q. Yuan,
\arXivid{0903.0122} [hep-ph];\\ A. Ibarra, A. Ringwald,
D. Tran and C. Weniger, \arXivid{0903.3625} [hep-ph]
\bibitem{ansol}
A. A. Andrianov and R. Soldati, \prd{51}{1995}{5961};\ \plb{435}{1998}{449};\\  A. A. Andrianov, R. Soldati and L. Sorbo, \prd{59}{1999}{025002}.
\bibitem{bumble} V.A. Kosteleck\'y and S. Samuel, \prd{39}{1989}{683};\\ V.A. Kosteleck\'y
and R. Potting, \npb{359}{1991}{545};\\ D. Colladay and V. A. Kostelecky,
\prd{58}{1998}{116002};\\ R. Bluhm and V.A. Kosteleck\'y, \prd{71}{2005}{065008};\\ J.L. Chkareuli, C.D. Froggatt and H.B. Nielsen, \prl{87}{2001}{091601};\ \npb{609}{2001}{46} .
\bibitem{guendelman} E. Guendelman and D.A. Owen, \plb{251}{1990}{439};\\ E. Strakhov and D.A. Owen, \jphg{22}{1996}{473} .
\bibitem{ae} A.A. Andrianov and D. Espriu, \plb{663}{2008}{450};\\ A.A. Andrianov, D. Espriu and V.A. Andrianov, \plb{678}{2009}{416} .
\bibitem{carroll} S.M. Carroll, G.B. Field, R. Jackiw, \prd{41}{1990}{1231}.
\bibitem{liveff1} D. Colladay and V.A. Kosteleck\'y, \prd{55}{1997}{6760};\  \prd{58}{1998}{116002};\\   S. Coleman and S. L. Glashow,
\prd{59}{1999}{116008}.
\bibitem{ags02}
A.A. Andrianov, P. Giacconi and R. Soldati,
\jhep{02}{2002}{030}.
\bibitem{adam} C. Adam and F. R. Klinkhamer,
\npb{607}{2001}{247};\\ V.A. Kosteleck\'y and R. Lehnert,
\prd{63}{2001}{065008};\\ R. Lehnert, \prd{68}{2003}{085003};\\
\bibitem{liveff2} C.D. Froggatt and H.B. Nielsen, {\it Origin of Symmetries},
World Scientific 1991;\\ G. M. Shore, \npb{717}{2005}{86};\\
T. Jacobson, S. Liberati and D. Mattingly, \ap{321}{2006}{150} ;\\
R. Montemayor and L.F. Urrutia,
\prd{72}{2005}{045018};\\  L. F. Urrutia,
\newjournal{Lect. Notes Phys.}{LNPHA}{702}{2006}{299};\\
V.A. Kostelecky and N. Russell,
  \arXivid{0801.0287} [hep-ph] .
\bibitem{kharzeev} D. Kharzeev, R.D. Pisarski, M.H.G. Tytgat, \prl{81}{1998}{512};\\
D. Kharzeev, \plb{633}{2006}{260} ;\\
D. Kharzeev, A. Zhitnitsky Nucl.Phys.A \npa{797}{2007}{67};\\
D.E. Kharzeev, L.D. McLerran, H.J. Warringa, \npa{803}{2008}{227};\\ D.E. Kharzeev, \arXivid{0906.2808} [hep-ph]  .
\bibitem{KSVZ} J.E. Kim, \prl{43}{1979}{103};\\
M.A. Shifman, A.I. Vainshtein and V.I. Zakharov, \npb{166}{1980}{493} .
\bibitem{aacgs06}
J. Alfaro, A.A. Andrianov, M. Cambiaso, P. Giacconi  and R. Soldati,
\plb{639}{2006}{586};\quad
\arXivid{0904.3557} [hep-th] (2009).
\bibitem{fan} Yi-Zhong Fan, \arXivid{0905.0908} [astro-ph.CO].
\bibitem{klink} C. Kaufhold and F.R. Klinkhamer,
\npb{734}{2006}{1}.
\bibitem{cmb} \href{http://www.gsi.de/fair/experiments/CBM/index_e.html}{Compressed Baryon Matter at FAIR} .
\bibitem{nica} \href{http://theor.jinr.ru/twiki/pub/NICA/WebHome/Wh_Paper_dk6.pdf}{NICA White Paper} (Eds. D. Blaschke, D. Kharzeev, A. Sissakian, A. Sorin, O. Teryaev, V. Toneev, I. Tserruya) .
\bibitem{axion} Hannestad S, Mirizzi A and Raffelt G, \newjournal{JCAP}{JCAPA}{0507}{2005}{002};\\
M. Kuster, G. Raffelt, B. Beltr\'an (Eds.),{\it Axions: Theory, Cosmology,
and Experimental Searches},\newjournal{Lect. Notes Phys.}{LNPHA}{2008}{741}{1}, Springer 2008 .
\bibitem{helio} K. Zioutas, M. Tsagri, T. Papaevangelou and T. Dafni,
\arXivid{0903.1807} [astro-ph.SR].
\bibitem{pvlas} E.~Zavattini {\it et al.}  [PVLAS Collaboration], \prd{77}{2008}{032006} .
\bibitem{cast} Zioutas K., et al. [CAST Collaboration], \prl{94}{2005}{121301};\\
E.~Arik {\it et al.}  [CAST Collaboration],
\newjournal{JCAP}{JCAPA}{0902}{2009}{008} .
\bibitem{massax} Y. Nomura and J. Thaler,
  \arXivid{0810.5397} [hep-ph].
\bibitem{PDG}  C.Amsler {\it et al.}  [Particle Data Group],
\plb{667}{2008}{1} .
\end{thebibliography}
\end{document}